%% file: IEEE-TAP-latex-template.tex
\documentclass[journal,9pt]{IEEEtran}

\usepackage{amsmath,amsfonts}
\usepackage{algorithmic}
\usepackage{algorithm}
\usepackage{array}
\usepackage{multirow} 
\usepackage{siunitx}
\usepackage{subfiles}
\usepackage[caption=false,font=normalsize,labelfont=sf,textfont=sf]{subfig}

\usepackage{textcomp}
\usepackage{stfloats}
\usepackage{url}
\usepackage{verbatim}
\usepackage{graphicx}
\usepackage{cite}
\usepackage{hyperref}
\usepackage{tikz}
\usepackage{pgfplots}
\usepackage{booktabs}

\usepackage{amsfonts}
\usepackage{pifont}
\usepackage{amsmath}
\usepackage{amssymb}
\usepackage{cancel}
\usepackage{xcolor}
\usepackage[english]{babel}
\usepackage[utf8]{inputenc}
\usepackage{listings}
\usepackage{import}

\usetikzlibrary{spy}
\usetikzlibrary{positioning, arrows.meta, shapes.geometric}
\pgfplotsset{compat=1.17}

\begin{document}

\title{Machine Learning-Driven Microwave Imaging for Soil Moisture Estimation near Leaky Pipe}
\author{\Large{Mohammad Ramezaninia, Mohammadreza Shams, Mohammad Zoofaghari}

\thanks{Mohammad Ramezaninia is with the Department of Electrical Engineering, Yazd University, Yazd 89195-741, Iran. email: m.ramezaninia@stu.yazd.ac.ir}
\thanks{Mohammadreza Shams is with the Mathematical Sciences Department of the Sharif
University of Technology, Tehran, Iran. (mohammadreza.shams95@student.sharif.edu).}
\thanks{Mohammad Zoofaghari is with the	Department of Electrical Engineering, Yazd University, Yazd 89195-741, Iran, and the Intervention Centre, Oslo University Hospital, 0372 Oslo, Norway. e-mail: zoofaghari@yazd.ac.ir}
}

\maketitle

\begin{abstract}
Characterizing soil moisture (SM) around drip irrigation pipes is crucial for precise and optimized farming. Machine learning (ML)  approaches are particularly suitable for this task as they can reduce uncertainties caused by soil conditions and the drip pipe positions, using features extracted from relevant datasets. This letter addresses local moisture detection in the vicinity of dripping pipes using a portable microwave imaging system. The employed ML approach is fed with two dimensional images generated by two different microwave imaging techniques based on spatio-temporal measurements at various frequency bands. The study investigates the performance of K-Nearest Neighbor (KNN) and Convolutional Neural
Networks (CNN) algorithms for moisture classification based on these images, both before and after performing soil clutter reduction. We also explore the potentials of CNN and KNN for moisture estimation around the plant roots and in the presence of pebbles.  The results demonstrate the more accurate moisture estimation using CNN when it is applied after clutter reduction considering back projection algorithm (BPA) as the imaging technique.    
\end{abstract}


\section{Introduction}\label{introduction}

Controlled leaky pipes are of  importance in mechanized agriculture for efficient irrigation, established to prevent water wastage and optimize plant growth. Accurately estimating moisture levels near plant roots provides valuable data for managing Subsurface Drip Irrigation (SDI) effectively. Consequently, for applications like field monitoring and plant growth assessment, frequent Soil Moisture (SM) observations are necessary.
SM estimation can be approached through model-based and data-driven strategies. The model-based approach offers speed and insights into critical factors like topography and vegetation, yet it often introduces uncertainties due to environmental manipulations. Conversely, data-driven methods demand extensive datasets for reliable results. These approaches typically leverage various Machine Learning (ML) techniques, including deep learning that consider different features of backscattered signals, including reflectivity and polarization to analyze and quantify SM content. The machines are trained through two major datasets from SM Active Passive (SMAP) and Cyclone Global Navigation Satellite System (CYGNSS) constellation in many studies on SM estimation \cite{roberts2022deep}. 
In this regard, the vegetation layer appears as a major obstacle that could be mitigated by the Water Cloud Model (WCL) and its modified version. 

In \cite{10237270}, the performance of various boosting algorithms is compared for SM estimation in vegetated areas considering WCL. This study indicates that categorical boosting algorithms outperform other methods when evaluated using Root Mean Square Error (RMSE) and mean relative error for moisture prediction. 
Convolutional Neural Network (CNN) is utilized in \cite{bagherkhani2024convolutional} for solving the inverse scattering problem of soil volumetric water content estimation. RMSE for the CNN model in this paper for the soil permittivity prediction is $0.1$. Furthermore, in \cite{9751623}, a deep Artificial Neural Network (ANN) model is utilized to retrieve SM based on training data generated from finite element simulations.
Also, prediction and forecasting of the root zone SM studied in \cite{carranza2022predicting} considering the random forest technique where the $RMSE=0.021$ is obtained. It is mentioned in this paper that random forests overestimate and underestimate the very low and very high moisture values respectively. 
In \cite{sun2022series}, a combinative technique is proposed to enhance deep learning by incorporating features from physical models. This study examines scenarios where the physical model  is applied both before and after the ML process, utilizing random forest , extreme gradient boosting , and light gradient boosting machine as the ML approaches. The results demonstrate that the combined physical model-ML approach outperforms the physical model and ML techniques when used separately.

Satellite data suffers from limited temporal and spatial resolutions, typically focusing on surface soil layers (e.g., depths up to 10 cm).
In some cases, point-based techniques using Ground Penetrating RADAR (GPR) or other sensors outperform remote sensing approaches due to their higher resolution, which is particularly beneficial for local moisture retrieval.   Reichman et al. \cite{reichman2017some}   investigate the performance evaluation of CNNs for detecting buried threats in images provided by GPR. The paper addresses two key challenges associated with applying CNNs to GPR data: the limited availability of training data and the determination of the optimal architecture.
Furthermore, \cite{zheng2019convolutional} 
investigates the water content classification using CNN based on GPR images. 
They structure the network into two components: one for feature selection and the other for classification and regression. 
Also, authors in \cite{Zhou} propose a novel approach for quantitative subsurface imaging using near-field scanning microwave microscopy with coaxial resonators, termed the Learning-Based (LB) method. The resolution enhancement by this method is confirmed by both numerical simulations and experimental data. This LB technique significantly reduces the runtime compared to conventional objective function methods as well.  
In \cite{zheng2019convolutional}  long short-term memory (LSTM) is exploited as a deep learning algorithm to create virtual soil moisture sensor through the data provided by a network of transducers such as air and soil temperature and humidity as well as solar radiation sensor. 

In this letter, we investigate a data-driven approach for the dataset of laboratory setup that we recently provided by an implemented Microwave Imaging System (MIS)  \cite{ramezaninia2024microwave}.  Here,  we explore the potential of Convolutional Neural Network (CNN) and K-Nearest Neighbors (KNN) to estimate SM around plant roots in the presence of pebbles instead of the model-based approach given by \cite{ramezaninia2024microwave}. We evaluate the performance of CNN and KNN using raw and processed images before and after clutter reduction.
Therefore, the main contributions of this paper are summarized as follows:
\begin{itemize}

\item Data-driven SM estimation using backscattered data from a ground-based portable system.
\item Establishing the impact of using a pre-processed image dataset for ML approaches.
\item Comparing the performance of two ML techniques for the current application.
\item Considering  the effect  of plant roots and soil pebbles in SM classification for SDI application.
  
\end{itemize}

\color{black}
In the following, we explore the image formation techniques used to create the dataset for our ML application. Section \label{proposed} details the fundamentals of CNN and KNN. Section \ref{Senarios} describes the laboratory setup for data acquisition, including the scenarios used for data generation. It also covers model development based on the provided dataset. Additionally, this section explores SM characterization for real dripping pipes. Finally, the paper concludes in Section \ref{conclusion}.

\section{Proposed approach}\label{proposed}

\begin{figure}[tb]
    \centering
    \begin{tikzpicture}[node distance=0.75cm, every node/.style={align=center, font=\small}, >=Stealth]
        \node (process2) [rectangle, minimum height=1.1cm, text width=1.9cm, fill=green!50!cyan!70] {Moist soil bag\\embedding};
        \node (process1) [rectangle, minimum height=1.1cm, text width=1.9cm, fill=blue!30!cyan!70, anchor=east, xshift=-1cm, yshift=2cm] at (process2.west) {System setup\\(n=1)};
        \node (process23) [rectangle, minimum height=1.1cm, text width=2.1cm, right=of process2, fill=red!50!yellow!80] {Data acquisition};
        \node (process233) [rectangle, minimum height=1.1cm, text width=1.9cm, above=of process23, fill=orange!70!brown!80] {Frequency setting};
        \node (process31) [rectangle, minimum height=1.1cm, text width=2.1cm, below=of process23, fill=orange!50!yellow!80] {Clutter reduction};
        \node (process3) [rectangle, minimum height=1.1cm, text width=2.1cm, left=of process31, fill=orange!40!brown!70] {Image formation (BAA and BPA)};
        \node (process4) [rectangle, minimum height=1.1cm, text width=2.1cm, below=of process3, fill=yellow!50!brown!70, yshift=-0.4cm] {$n = n + 1$};
        \node (process5) [diamond, minimum height=0.4cm, text width=1.2cm, left=of process4, fill=purple!40!blue!70, xshift=-0.2cm] {$n>n_s$};
        \node (process6) [rectangle, minimum height=1.1cm, text width=2.1cm, below=of process5, fill=orange!50!brown!70, yshift=-0.5cm] {Machine training};
        \draw [->, line width=1pt] (process6.east) -- ++(2cm, 0) node[midway, above, draw=none] {SM model};
        \draw [->, line width=1pt] (process2) -- (process23);
        \draw [->, line width=1pt] (process233) -- (process23);
        \draw [->, line width=1pt] (process23) -- (process31);
        \draw [->, line width=1pt] (process31) -- (process3);
        \draw [->, line width=1pt] (process3) -- (process4);
        \draw [->, line width=1pt] (process4) -- (process5);
        \draw [->, line width=1pt, looseness=2, out=90] (process5) -- node[anchor=east, near start, draw=none] {YES} (process6);
        \draw [->, line width=1pt] (process5.north) -- ++(0,0) |- (process2.west) node[pos=0.25, anchor=south east, draw=none] {NO};
        \draw [-, line width=1pt] (process1.south) -- ++(0, -1cm) -| (process5.north);
    \end{tikzpicture}
    \caption{Flowchart of proposed approach for SM model extraction using $n_s$ wet soil bags with known SM contents.}
    \label{main_flowchart}
\end{figure}
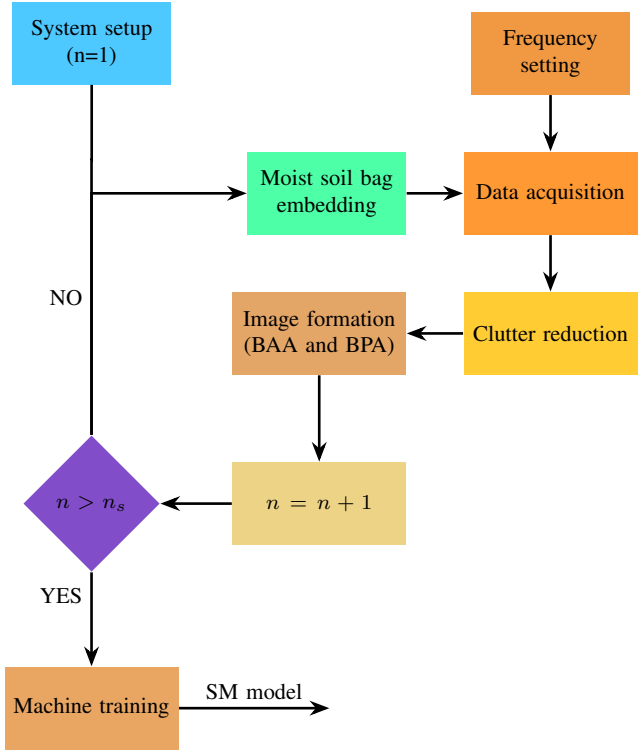

\color{black}
In this study we use the image data for the SM retrieval. These images are significantly influenced by varying levels of moisture content, resulting in the amplification of moist areas and the deterioration of pipe reflections. The captured subsurface images around the pipelines, highlighting specific moisture areas, are used to train the ML algorithms and  classify the moisture content around a dripping pipe  as outlined in Fig. \ref{main_flowchart}. 
  In this process, spatio-temporal data is collected after placing artificial soil bags with varying moisture levels over the pipeline. Utilizing a clutter reduction and a microwave imaging technique, images are generated at different frequency intervals to serve as input for the ML model. 
\color{black}

\subsection{Image formation}\label{image formation}
In this letter, the images required for training the learners are constructed using the Born Approximation (BAA) and Back Projection (BPA) algorithms, both of which are well-known in the literature.
 BAA is represented by an integral equation involving equivalent contrast sources across each mesh in the reconstruction domain. The measured data, collected by a vector network analyzer along a scan line, consists of backscattered step frequency continuous waves. This data is then transformed into a matrix equation, which is solved for the unknown contrast sources using a regularization technique such as the truncated Singular Value Decomposition (SVD) method.
 
Additionally, BPA compensates for the phase of the backscattered field due to the round-trip signal path between the antenna and each pixel in the imaging area. This process is carried out for all frequencies and antenna positions (B-scan data), achieving synthetic focusing over each pixel of the reconstructed image where the scatterer appears.
The formulation of BAA and BPA are detailed in \cite{ramezaninia2024microwave} for current application. 
Also, clutter reduction is achieved using the SVD technique. In this method, the clutter effect is minimized by removing one or two singular vectors associated with the largest singular values of the B-scan matrix.

\color{black}
\subsection{ML techniques}

\begin{figure*}[htb]
\centering
\includegraphics[scale=0.3]{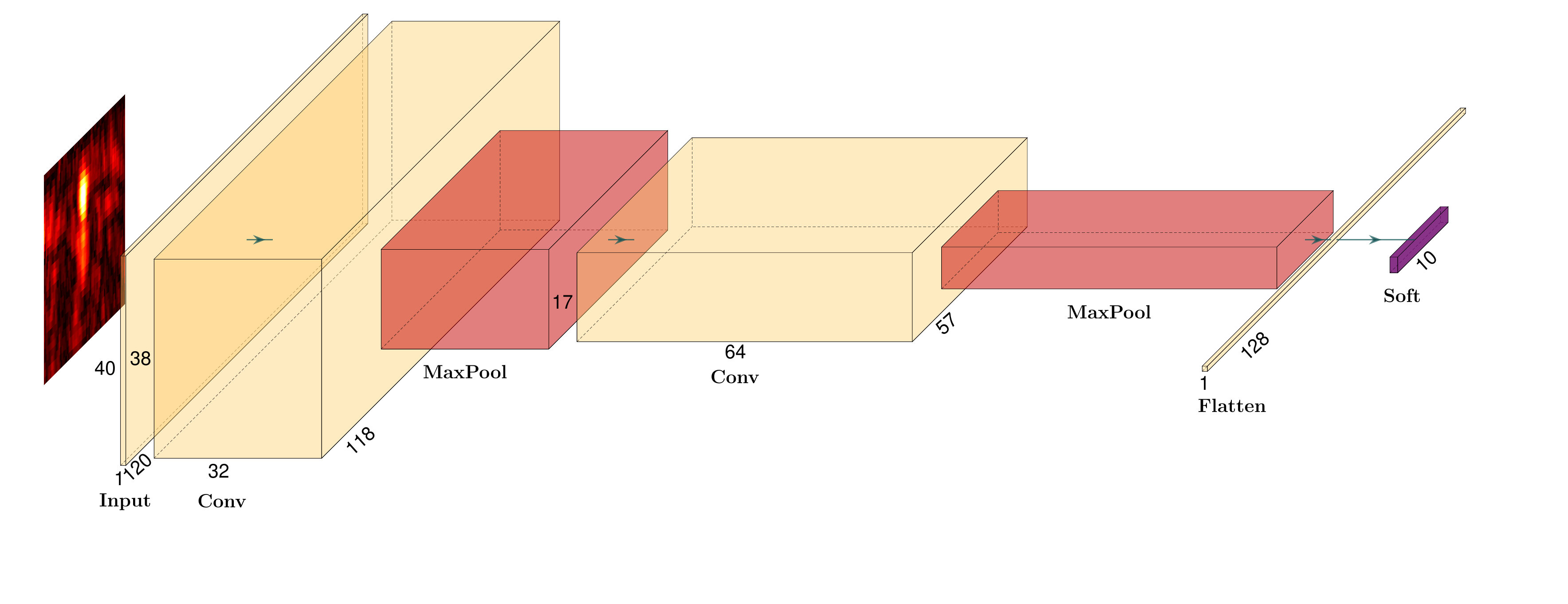}
\caption{\small The architecture of the convolutional neural network.}
\label{Soil_moisture_estimation_with_AI}
\end{figure*}

CNNs are specialized type of neural network architectures well-suited for processing grid-like data such as images or matrices. These models consist of one or more convolutional layers that learn to extract relevant features from the input. Within each convolutional layer, a set of learnable filters are applied across the input, performing element-wise multiplication and summation to produce a feature map. This allows the network to detect local patterns and learn spatial relationships in the data. To manage the dimensionality of the feature maps and capture higher-level abstractions, pooling layers are often inserted after the convolutional layers. The pooling operation aggregates the feature map values within a local neighborhood, reducing the spatial size and introducing translation invariance. Finally, the flattening layer converts the 2D feature maps into a 1D feature vector that can be fed into a fully-connected layer for classification or regression tasks. The specific CNN architecture used in this study is depicted in Fig. \ref{Soil_moisture_estimation_with_AI}, with the number of convolutional layers carefully chosen to balance accurate classification performance and computational efficiency.

\begin{figure}[htb]
\centering
\includegraphics[scale=0.7]{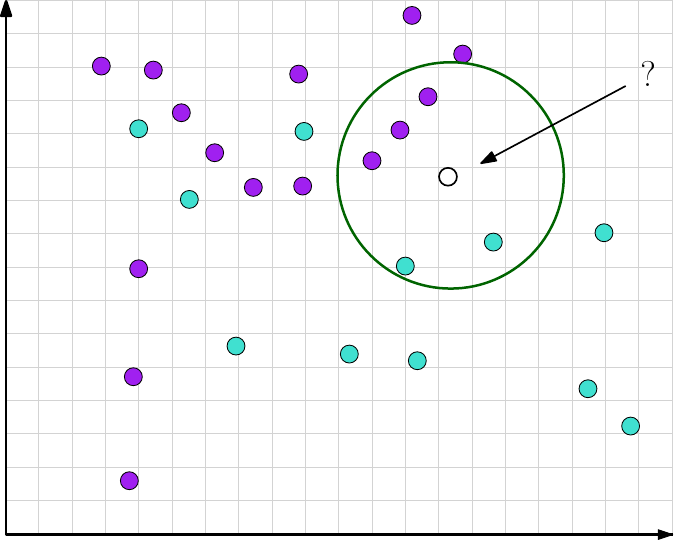}
\caption{\small  A practical example of KNN classification.}
\label{KNN_example}
\end{figure}

KNN algorithm is a simple yet effective ML model for classification tasks. It works by identifying the K closest data points to a given test instance and assigning the most common label among those neighbors to the test instance. More specifically, the KNN model first transforms all data into a Euclidean space, where the distance between data points can be calculated. For a test instance, the algorithm determines the distances between the test point and all training data points. It then selects the K training points that are closest to the test point and assigns the label that occurs most frequently among those K neighbors to the test instance. For example, as illustrated in Fig. \ref{KNN_example}, if K=5, to predict the label of the white circle, the algorithm would identify the five nearest neighbors and assign the label that is most common among those five points (in this case, purple) to the white circle.

\section{Scenarios \& results}\label{Senarios}

In this section, to evaluate the proposed method for localized moisture estimation, data acquisition tool is introduced and  the machine training and evaluation  as well as moisture classification around a  dripping laboratory buried pipe is discussed.  Also, we assess the performance of the suggested technique in the presence of interfering objects like pebbles and plant roots.

\subsection{Laboratory setup}
\begin{figure}[htb]
\centering
\includegraphics[scale=0.25]{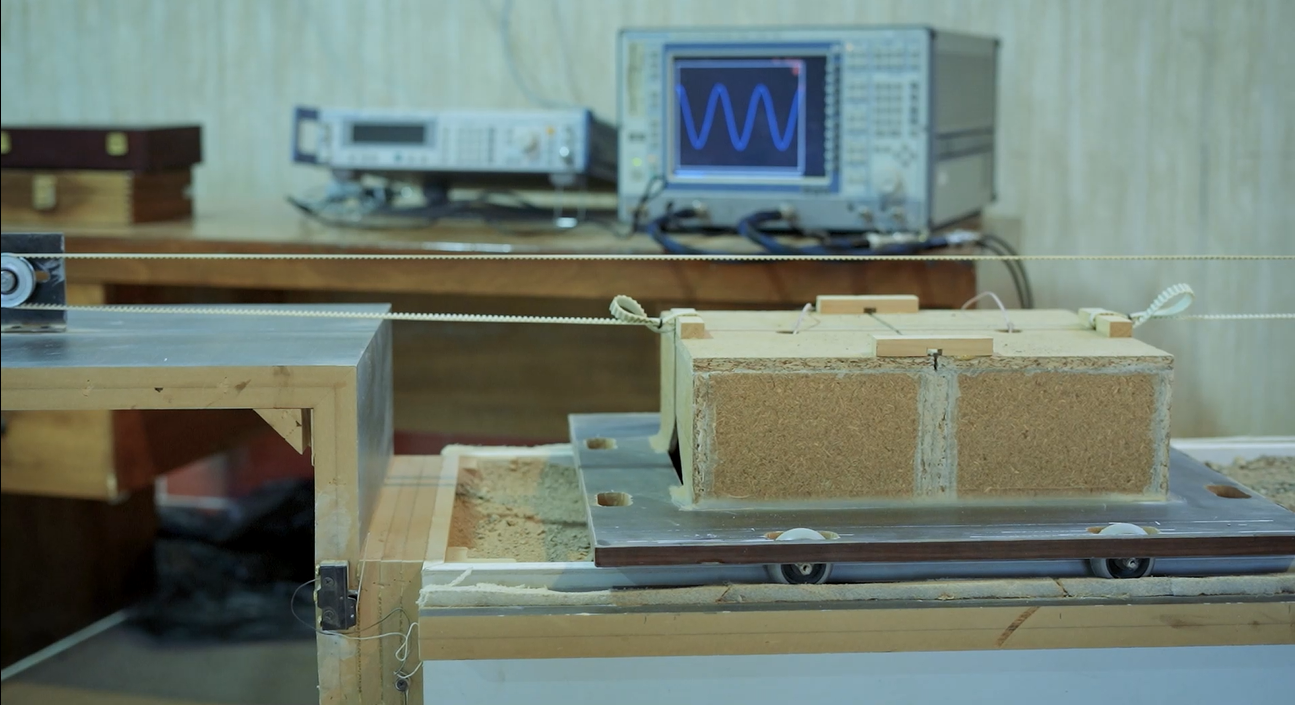}
\caption{\small A view of implemented laboratory setup for SM estimation.}
\label{one_pipe2}
\end{figure}

Our testbed, which includes bare soil, is constructed using a wooden box measuring 40 cm × 60 cm × 120 cm. This setup contains a buried PVC pipe analogous to a drip irrigation system. The two ends of the pipe are cap-sealed and waterproofed, with two thin hoses connected for water inlet and outlet.
Data acquisition is carried out using two bent dipole antennas  serving as the transmitting and receiving antennas. The transmit path includes a 21 dB-gain amplifier, and an 8 GHz vector network analyzer used for Step Frequency Continuous Wave (SFCW)  generation and reception. The system parameters are listed in Tab. \ref{tab:mytable}. Data collection process is automated by mounting the antennas on four wheels and connecting the antenna box to a stepper motor  controlled through an Arduino microcontroller. Communication between the computer and Arduino is established to synchronize the data acquisition and antenna movement, as illustrated in Fig. \ref{one_pipe2}. 

.


	

\begin{table}[htbp]
	\centering
	\caption{Data acquisition parameters for reference scenario}
	\label{tab:mytable}
	\begin{tabular}{cccc}
		\toprule
		Parameters & Symbol & Value & Unit \\
		\midrule
		Transmitter power&  $P_t$ &  $15$ & $dBm$ \\
		 Length of scan line &  $L$ &  $1.2$ &  $m$ \\
		 Frequency band &  $f$ &  $1.2-3.775$ &  $GHz$ \\
		 Frequency step &  $\Delta f$ &  $25$ &  $MHz$ \\
		 No. data acquisition points &  $N_s$ &  $45$ &   \\
		 Tx \& Rx antenna Gain &  $G_t \& G_r$ &  $7$ &  $dB$ \\
		 LNA gain      &  $G_{LNA}$ &  $21$ &  $dB$ \\
		 Scan duration &  $T$     &  $14$ &  $min$ \\
          Pipe diameter & $D$    &   $4.5$ & $cm$ \\
          Pipe depth    & $d$ &$12$   &$cm$\\
          Space between frequency bands&$\delta f$&$10$&$MHz$\\
		\bottomrule
	\end{tabular}
\end{table}
\subsection{Model development and evaluation }\label{model_gene}
We reconstructed BAA and BPA images for eight moist soil bags with moisture content ranging from $12.5\%$ to $100\%$, placed on a water-carrying pipe. We investigated nine scenarios outlined in Tab. \ref{senarios}, where each scenario differs from the reference scenario (scenario 3) based on specified parameters. For each soil bag, we captured sixteen images across sixteen frequency bands spaced by 10 MHz, resulting in a dataset of 128 elements. This dataset was then used to train both CNN and KNN models.
\begin{table*}[htbp]
\centering
\caption{Accuracy of proposed ML models for different scenarios of data  acquisition and processing}
\label{senarios}
\begin{tabular}{ccccc}
\toprule
Scenario No. & Description & Accuracy(CNN) & Accuracy(KNN) &  \\ \midrule
1  & Raw data & $69\%$ & $24\%$&  \\
2  & Clutter-reduced data  &$89\%$ & $58\%$& \\
3  & Reference Scenario& $100\%$(BPA)-$100\%$(BAA) & $100\%$(BPA)-$100\%$(BAA) &    \\
4  & Eight lower  frequency bands &  $100\%$(BPA)-$100\%$(BAA) & $93\%$(BPA)-$100\%$(BAA) &   \\
5  & Eight higher frequency bands & $100\%$(BPA)-$100\%$(BAA) & $100\%$(BPA)-$100\%$(BAA)  &  \\
6  & $\Delta f=75MHz$ &  $100\%$(BPA)-$100\%$(BAA) & $93\%$(BPA)-$93\%$(BAA)  & \\
7  & $N_s=23$  & $100\%$(BPA)-$100\%$(BAA) & $100\%$(BPA)-$100\%$(BAA) & \\
8 & $L=0.6$m & $100\%$(BPA)-$100\%$(BAA) & $96\%$(BPA)-$93\%$(BAA) &   \\
9 & $\delta f =50 MHz$ & $93\%$(BPA)-$100\%$(BAA) & $86\%$(BPA)-$93\%$(BAA) &    \\
\bottomrule
\end{tabular}
\end{table*}

In each scenario, we allocated $60\%$ of the data for model training, $20\%$ for model validation, and $20\%$ for model testing. The first scenario involved raw B-scan data collected by the setup, the second scenario used pre-processed data with clutter reduction, and the third scenario utilized image data created by BAA and BPA.
As indicated in the Tab. \ref{senarios} , both CNN and KNN models demonstrated lower accuracy before clutter reduction (raw data) and in both scenarios CNN outperforms KNN.  After applying BAA and BPA, the moisture estimation accuracy improved by approximately $10\%$ for CNN and $40\%$ for KNN (reference scenario).  Scenarios 4 to 9 investigated the impact of various system parameters, such as frequency band, frequency step, number of data acquisition points, and scan line length. Tab. \ref{senarios} specifically shows that the CNN algorithm demonstrates high accuracy across various scenarios when using the BPA image dataset. The results from these scenarios showed that both CNN and KNN are highly robust against variations in data acquisition parameters.

\begin{figure*}[tb]
	\centering
 
 	\subfloat[]{\input{figures/CNF_MAT_for_RAW_data_bef_clut_red.tex}}
	\hfill
	\subfloat[]{\input{figures/CNF_MAT_raw_aft_clut_red.tex}}
	\hfill
	\subfloat[]{\input{figures/CNF_MAT_back_pro_f_1_2_16.tex}}
	\caption{ Confusion matrix for SM estimation by CNN with (a) raw data, (b) after clutter reduction and  (c) after image formation.}
	\label{conf_mat}
\end{figure*}
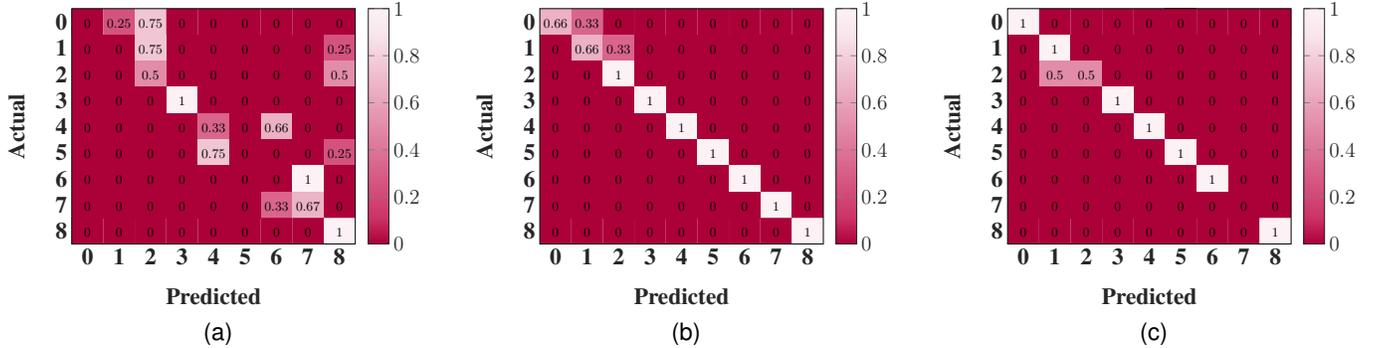

 Fig. \ref{conf_mat} displays the confusion matrix for CNN model. As observed, the model accuracy deteriorates for both high and low moisture level considering raw B-scan data.   Additionally, after clutter removal, the accuracy of model  significantly increases specially for high moisture contents. Due to more sensitivity of images to the moisture changes. Fig. \ref{conf_mat} also depict the confusion matrix after image formation using BPA denoting the accuracy improvement even for low moisture levels retrieval.

\subsection{SM estimation in the presence of medium clutter}
Tab. \ref{Senorios_with_clut} illustrates the estimated soil moisture by the aforementioned ML algorithms in the presence of soil pebbles and plant roots at various densities for both dry soil and soil with $60\%$ moisture content. As shown, both CNN and KNN overestimate the SM for dry soil and underestimate SM for wet soil in the presence of soil clutters. The CNN results are more accurate, especially for low clutter density.

\begin{table}
\centering
\caption{Accuracy of the proposed algorithms in the presence of various  plant roots and soil pebbles densities}
\label{Senorios_with_clut}
\begin{tabular}{ccccccc}
\toprule
\multirow{2}{*}{SM} & \multirow{2}{*}{Clutter Type} & \multirow{2}{*}{Density} &\multicolumn{2}{c}{CNN} & \multicolumn{2}{c}{KNN} \\ \cmidrule{4-7}
 &  &  & BPA&BAA & BPA & BAA \\ \midrule
0\%&Root &Low & 8\%  &  7.8\%   &  40\% & 12.5\%  \\
0\% & Root &High & 25\% & 25\% & 35\% & 3\% \\
 0\%  &  Gravel &Low & 25\% & 25\% & 18\% & 3\%\\
 0\%  &  Gravel &High  & 25\% & 25\% & 32\% & 23\% \\
 60\% &  Root &Moderate & 50\% & 50\% & 50\% &  80\%\\
 60\% &  Gravel &Moderate & 50\% & 50\% & 50\% & 53\% \\
\bottomrule
\end{tabular}
\end{table}

\subsection{Laboratory SDI result}
In this subsection, we investigate SM estimation for a real leakage from a punctured pipe. We captured eleven B-scans, each within fourteen minutes. For each scan, we obtained sixteen different images corresponding to sixteen frequency bands. We applied the model developed in Section \ref{model_gene} to each image given by a frequency band and reported the SM as the average of values from the sixteen frequency bands. Figs. \ref{real_leak_CNN} and \ref{real_leak_KNN} illustrate the estimated SMs using BPA and BAA as the image formation methods provided by CNN and KNN models versus time, respectively. In all cases, clutter reduction was performed by removing the singular vector of the B-scan matrix associated with the largest singular value.
The results clearly demonstrate that using BPA along with the CNN model yields the best match with real moisture measurements using the gravimetric method. This outcome aligns with the high precision of the CNN model and the superior performance of the BPA algorithm compared to BAA, as evidenced in our previous research \cite{ramezaninia2024microwave}.

\begin{figure}[tb]
	\centering
	\subfloat[]{\input{figures/plot_one.tex}}
	
	\subfloat[]{\input{figures/CNN_Born_real_leak.tex}}
	
	\caption{SM estimation for a real leakage considering (a) BPA and (b) BAA for the image formation in CNN model.}
	\label{real_leak_CNN}
\end{figure}

\begin{figure}[tb]
	\centering
	\subfloat[]{\input{figures/plot_BP_KNN.tex}}

	\subfloat[]{\input{figures/KNN_real_leakage_BOrn.tex}}
	
	\caption{SM estimation for a real leakage considering (a) BPA and (b) BAA for the image formation in KNN model.}
	\label{real_leak_KNN}
\end{figure}

\section{Conclusion }\label{conclusion}

This letter explores the potential of machine learning techniques for estimating soil moisture in a ground-based laboratory setup. We provided a dataset containing spatio-temporal data captured at various frequency bands for several embedded moist soil bags. This dataset was used to train two well-known deep learning methods: Convolutional Neural Network (CNN) and K-Nearest Neighbors (KNN). The proposed classification model, validated by test data, is applied to predict the moisture content of dripping pipes.
Our results demonstrate that preprocessing the data through clutter reduction enhances the accuracy of CNN and KNN by up to $20\%$ and $30\%$, respectively. Furthermore, the application of microwave imaging techniques improves accuracy by up to $10\%$ for CNN and $40\%$ for KNN. The findings also indicate that CNN outperforms KNN in predicting soil moisture around real dripping pipes, even in the presence of plant roots and soil pebbles.
The outcomes of this study suggest that machine learning techniques, especially CNN, are highly effective for local soil moisture estimation in complex environments. This advancement could be leveraged to develop mechanistic irrigation systems, optimizing water usage and enhancing agricultural productivity.
\color{black}


\bibliographystyle{IEEEtran}
\bibliography{IEEE-TAP-latex-template}

\vfill

\end{document}

%% file: figures/CNF_MAT_for_RAW_data_bef_clut_red.tex
\begin{tikzpicture}[scale = 0.55]
 \begin{axis}[
   colormap={purplewhite}{
       color=(purple!90!black);
       color=(purple!80);
       color=(purple!70);
       color=(purple!60);
       color=(purple!50);
       color=(purple!40);
       color=(purple!30);
       color=(purple!20);
       color=(purple!10);
       color=(purple!5)},
   xlabel=Predicted,
   xlabel style={font=\bfseries\fontsize{16}{16}\selectfont, color=white!15!black},
   xlabel style={yshift=-10pt},
   ylabel=Actual,
   ylabel style={font=\bfseries\fontsize{16}{16}\selectfont, color=white!15!black},
   ylabel style={yshift=15pt},
   xticklabels={0, 1, 2, 3, 4, 5, 6, 7, 8}, 
   xticklabel style={font=\bfseries\fontsize{16}{16}\selectfont, color=white!15!black},
   xtick={0,...,8}, 
   xtick style={draw=none},
   yticklabels={0, 1, 2, 3, 4, 5, 6, 7, 8}, 
   yticklabel style={font=\bfseries\fontsize{16}{16}\selectfont, color=white!15!black},
   ytick={0,...,8}, 
   ytick style={draw=none},
   enlargelimits=false,
   colorbar,
   colorbar style={
       font=\bfseries\fontsize{14}{14}\selectfont, color=white!15!black},
   nodes near coords={\pgfmathprintnumber[precision=2]\pgfplotspointmeta},
   nodes near coords style={
       yshift=-7pt
   },
  ]
  \addplot[
   matrix plot,
   mesh/cols=9, 
   point meta=explicit,
   draw opacity=0,
   fill
  ] table [meta=C] {
   x y C
   0 0 0.0
   1 0 0.25
   2 0 0.75
   3 0 0.0
   4 0 0.0
   5 0 0.0
   6 0 0.0
   7 0 0.0
   8 0 0.0
    
   0 1 0.0
   1 1 0.0
   2 1 0.75
   3 1 0.0
   4 1 0.0
   5 1 0.0
   6 1 0.0
   7 1 0.0
   8 1 0.25
    
   0 2 0.0
   1 2 0.0
   2 2 0.5
   3 2 0.0
   4 2 0.0
   5 2 0.0
   6 2 0.0
   7 2 0.0
   8 2 0.5
  
   0 3 0.0
   1 3 0.0
   2 3 0.0
   3 3 1.0
   4 3 0.0
   5 3 0.0
   6 3 0.0
   7 3 0.0
   8 3 0.0
  
   0 4 0.0
   1 4 0.0
   2 4 0.0
   3 4 0.0
   4 4 0.33
   5 4 0.0
   6 4 0.66
   7 4 0.0
   8 4 0.0

   0 5 0.0
   1 5 0.0
   2 5 0.0
   3 5 0.0
   4 5 0.75
   5 5 0.0
   6 5 0.0
   7 5 0.0
   8 5 0.25

   0 6 0.0
   1 6 0.0
   2 6 0.0
   3 6 0.0
   4 6 0.0
   5 6 0.0
   6 6 0
   7 6 1
   8 6 0.0

   0 7 0.0
   1 7 0.0
   2 7 0.0
   3 7 0.0
   4 7 0.0
   5 7 0.0
   6 7 0.33
   7 7 0.6666666666666666
   8 7 0.0

   0 8 0.0
   1 8 0.0
   2 8 0.0
   3 8 0.0
   4 8 0.0
   5 8 0.0
   6 8 0.0
   7 8 0.0
   8 8 1.0
  };
 \end{axis}
\end{tikzpicture}

%% file: figures/CNF_MAT_raw_aft_clut_red.tex
\begin{tikzpicture}[scale = 0.55]
 \begin{axis}[
   colormap={purplewhite}{
       color=(purple!90!black);
       color=(purple!80);
       color=(purple!70);
       color=(purple!60);
       color=(purple!50);
       color=(purple!40);
       color=(purple!30);
       color=(purple!20);
       color=(purple!10);
       color=(purple!5)},
   xlabel=Predicted,
   xlabel style={font=\bfseries\fontsize{16}{16}\selectfont, color=white!15!black},
   xlabel style={yshift=-10pt},
   ylabel=Actual,
   ylabel style={font=\bfseries\fontsize{16}{16}\selectfont, color=white!15!black},
   ylabel style={yshift=15pt},
   xticklabels={0, 1, 2, 3, 4, 5, 6, 7, 8}, 
   xticklabel style={font=\bfseries\fontsize{16}{16}\selectfont, color=white!15!black},
   xtick={0,...,100}, 
   xtick style={draw=none},
   yticklabels={0, 1,2 , 3, 4, 5, 6, 7, 8}, 
   yticklabel style={font=\bfseries\fontsize{16}{16}\selectfont, color=white!15!black},
   ytick={0,...,100}, 
   ytick style={draw=none},
   enlargelimits=false,
   colorbar,
      colorbar style={
font=\bfseries\fontsize{14}{14}\selectfont, color=white!15!black},
   nodes near coords={\pgfmathprintnumber[precision=2]\pgfplotspointmeta},
   nodes near coords style={
    yshift=-7pt
   },
  ]
  \addplot[
   matrix plot,
   mesh/cols=9, 
   point meta=explicit,
   draw opacity=0,
   fill
  ] table [meta=C] {
   x y C
   0 0 0.66
   1 0 0.33
   2 0 0
   3 0 0
   4 0 0
   5 0 0
   6 0 0
   7 0 0
   8 0 0
    
   0 1 0
   1 1 0.66
   2 1 0.33
   3 1 0
   4 1 0
   5 1 0
   6 1 0
   7 1 0
   8 1 0
    
   0 2 0
   1 2 0
   2 2 1
   3 2 0
   4 2 0
   5 2 0
   6 2 0
   7 2 0
   8 2 0
  
   0 3 0
   1 3 0
   2 3 0
   3 3 1
   4 3 0
   5 3 0
   6 3 0
   7 3 0
   8 3 0
  
   0 4 0
   1 4 0
   2 4 0
   3 4 0
   4 4 1
   5 4 0
   6 4 0
   7 4 0
   8 4 0

   0 5 0
   1 5 0
   2 5 0
   3 5 0
   4 5 0
   5 5 1
   6 5 0
   7 5 0
   8 5 0

   0 6 0
   1 6 0
   2 6 0
   3 6 0
   4 6 0
   5 6 0
   6 6 1
   7 6 0
   8 6 0

   0 7 0
   1 7 0
   2 7 0
   3 7 0
   4 7 0
   5 7 0
   6 7 0
   7 7 1
   8 7 0

   0 8 0
   1 8 0
   2 8 0
   3 8 0
   4 8 0
   5 8 0
   6 8 0
   7 8 0
   8 8 1
  };
 \end{axis}
\end{tikzpicture}

%% file: figures/CNF_MAT_back_pro_f_1_2_16.tex
\begin{tikzpicture}[scale = 0.55]
 \begin{axis}[
   colormap={purplewhite}{
       color=(purple!90!black);
       color=(purple!80);
       color=(purple!70);
       color=(purple!60);
       color=(purple!50);
       color=(purple!40);
       color=(purple!30);
       color=(purple!20);
       color=(purple!10);
       color=(purple!5)},
   xlabel=Predicted,
   xlabel style={font=\bfseries\fontsize{16}{16}\selectfont, color=white!15!black},
   xlabel style={yshift=-10pt},
   ylabel=Actual,
   ylabel style={font=\bfseries\fontsize{16}{16}\selectfont, color=white!15!black},
   ylabel style={yshift=15pt},
   xticklabels={0, 1, 2, 3, 4, 5, 6, 7, 8}, 
   xticklabel style={font=\bfseries\fontsize{16}{16}\selectfont, color=white!15!black},
   xtick={0,...,100}, 
   xtick style={draw=none},
   yticklabels={0, 1,2 , 3, 4, 5, 6, 7, 8}, 
   yticklabel style={font=\bfseries\fontsize{16}{16}\selectfont, color=white!15!black},
   ytick={0,...,100}, 
   ytick style={draw=none},
   enlargelimits=false,
   colorbar,
   colorbar style={
font=\bfseries\fontsize{14}{14}\selectfont, color=white!15!black},
   nodes near coords={\pgfmathprintnumber[precision=2]\pgfplotspointmeta},
   nodes near coords style={
    yshift=-7pt
   },
  ]
  \addplot[
   matrix plot,
   mesh/cols=9, 
   point meta=explicit,
   draw opacity=0,
   fill
  ] table [meta=C] {
   x y C
   0 0 1
   1 0 0
   2 0 0
   3 0 0
   4 0 0
   5 0 0
   6 0 0
   7 0 0
   8 0 0
    
   0 1 0
   1 1 1
   2 1 0
   3 1 0
   4 1 0
   5 1 0
   6 1 0
   7 1 0
   8 1 0
    
   0 2 0
   1 2 0.5
   2 2 0.5
   3 2 0
   4 2 0
   5 2 0
   6 2 0
   7 2 0
   8 2 0
  
   0 3 0
   1 3 0
   2 3 0
   3 3 1
   4 3 0
   5 3 0
   6 3 0
   7 3 0
   8 3 0
  
   0 4 0
   1 4 0
   2 4 0
   3 4 0
   4 4 1
   5 4 0
   6 4 0
   7 4 0
   8 4 0

   0 5 0
   1 5 0
   2 5 0
   3 5 0
   4 5 0
   5 5 1
   6 5 0
   7 5 0
   8 5 0

   0 6 0
   1 6 0
   2 6 0
   3 6 0
   4 6 0
   5 6 0
   6 6 1
   7 6 0
   8 6 0

   0 7 0
   1 7 0
   2 7 0
   3 7 0
   4 7 0
   5 7 0
   6 7 0
   7 7 0
   8 7 0

   0 8 0
   1 8 0
   2 8 0
   3 8 0
   4 8 0
   5 8 0
   6 8 0
   7 8 0
   8 8 1
  };
 \end{axis}
\end{tikzpicture}

%% file: figures/plot_one.tex
\begin{tikzpicture}[scale = 0.8]
  \begin{axis}[
    xlabel={Time[min]},
       xlabel style={font=\bfseries\fontsize{12}{12}\selectfont, color=white!15!black},
    ylabel={SM[\%]},
       ylabel style={font=\bfseries\fontsize{12}{12}\selectfont, color=white!15!black},
    ymin=0,
    ymax=100,
       xticklabel style={font=\bfseries\fontsize{12}{12}\selectfont, color=white!15!black},
          yticklabel style={font=\bfseries\fontsize{12}{12}\selectfont, color=white!15!black},
    legend pos=north west
  ]
    \addplot[green,mark=o] coordinates {
        (14, 33.59375)
        (28, 12.5)
        (42, 29.6875)
        (56, 29.6875)
        (70, 29.6875)
        (84, 12.5)
        (98, 25.0)
        (112, 19.53125)
        (126, 25.0)
        (140, 20.3125)
        (154, 20.3125)
    };
    \addlegendentry{ML-Estimated}
\
    \addplot[red] coordinates {
        (14, 26.8)
        (154, 19.8)
    };
    \addlegendentry{Linear Fitting}
    \addplot[blue,dashed] coordinates {(14,22) (154,22)};
    \addlegendentry{Measurement}
  \end{axis}
\end{tikzpicture}

%% file: figures/CNN_Born_real_leak.tex
\begin{tikzpicture}[scale = 0.8]
  \begin{axis}[
    xlabel={Time[min]},
       xlabel style={font=\bfseries\fontsize{12}{12}\selectfont, color=white!15!black},
    ylabel={SM[\%]},
       ylabel style={font=\bfseries\fontsize{12}{12}\selectfont, color=white!15!black},
    ymin=0,
    ymax=100,
       xticklabel style={font=\bfseries\fontsize{12}{12}\selectfont, color=white!15!black},
          yticklabel style={font=\bfseries\fontsize{12}{12}\selectfont, color=white!15!black},
    legend pos=north west
  ]
    \addplot[green,mark=o] coordinates {
        (14, 39.0625)
        (28, 32.8125)
        (42, 37.5)
        (56, 37.5)
        (70, 37.5)
        (84, 37.5)
        (98, 42.1875)
        (112, 42.1875)
        (126, 42.1875)
        (140, 47.65625)
        (154, 43.75)
    };
    \addlegendentry{ML-Estimated}

    \addplot[red] coordinates {
        (14, 34.9380)
        (154, 45.0180)
    };
    \addlegendentry{Linear Fitting}
    \addplot[blue,dashed] coordinates {(14,22) (154,22)};
    \addlegendentry{Measurement}
  \end{axis}
\end{tikzpicture}

%% file: figures/plot_BP_KNN.tex
\pgfplotsset{compat=1.17}
\begin{tikzpicture}[scale=0.8]
  \begin{axis}[
    xlabel={Time[min]},
       xlabel style={font=\bfseries\fontsize{12}{12}\selectfont, color=white!15!black},
    ylabel={SM[\%]},
       ylabel style={font=\bfseries\fontsize{12}{12}\selectfont, color=white!15!black},
    ymin=0,
    ymax=100,
       xticklabel style={font=\bfseries\fontsize{12}{12}\selectfont, color=white!15!black},
          yticklabel style={font=\bfseries\fontsize{12}{12}\selectfont, color=white!15!black},
    legend pos=north west
  ]
    \addplot[green,mark=o] coordinates {
        (14, 50.0)
        (28, 50.78125)
        (42, 49.21875)
        (56, 50.0)
        (70, 50.0)
        (84, 50.78125)
        (98, 56.25)
        (112, 40.625)
        (126, 53.125)
        (140, 55.46875)
        (154, 54.6875)
    };
    \addlegendentry{ML-Estimated}

    \addplot[red] coordinates {
        (14, 49.1080)
        (154, 52.888)
    };
    \addlegendentry{Linear Fitting}
    \addplot[blue,dashed] coordinates {(14,22) (154,22)};
    \addlegendentry{Measurement}
  \end{axis}
\end{tikzpicture}

%% file: figures/KNN_real_leakage_BOrn.tex
\begin{tikzpicture}[scale = 0.8]
  \begin{axis}[
    xlabel={Time[min]},
       xlabel style={font=\bfseries\fontsize{12}{12}\selectfont, color=white!15!black},
    ylabel={SM[\%]},
       ylabel style={font=\bfseries\fontsize{12}{12}\selectfont, color=white!15!black},
    ymin=0,
    ymax=100,
       xticklabel style={font=\bfseries\fontsize{12}{12}\selectfont, color=white!15!black},
          yticklabel style={font=\bfseries\fontsize{12}{12}\selectfont, color=white!15!black},
    legend pos=north west
  ]
    \addplot[green,mark=o] coordinates {
        (14, 47.65625)
        (28, 11.71875)
        (42, 16.40625)
        (56, 38.28125)
        (70, 35.15625)
        (84, 37.5)
        (98, 46.875)
        (112, 45.3125)
        (126, 39.0625)
        (140, 54.6875)
        (154, 50.0)
    };
    \addlegendentry{ML-Estimated}

    \addplot[red] coordinates {
        (14, 25.82)
        (154, 51.02)
    };
    \addlegendentry{Linear Fitting}
    \addplot[blue,dashed] coordinates {(14,22) (154,22)};
    \addlegendentry{Measurment}
  \end{axis}
\end{tikzpicture}